\documentclass[preprint]{elsarticle}

\usepackage{hyperref}
\usepackage{subfigure}
\usepackage{booktabs}

\usepackage{lineno,hyperref}

\journal{Journal of Nuclear Instruments and Methods in Physics Research}

\begin{document}

\begin{frontmatter}

\title{Photoelectron Yields of Scintillation Counters with Embedded Wavelength-Shifting Fibers Read Out 
With Silicon Photomultipliers}

\author[DUBNAaddress]{Akram Artikov}
\author[DUBNAaddress]{Vladimir Baranov}
\author[NIUaddress]{Gerald C. Blazey}
\author[UVAaddress]{Ningshun Chen}
\author[DUBNAaddress,TSUHEPIaddress]{Davit Chokheli}
\author[DUBNAaddress]{Yuri Davydov}
\author[UVAaddress]{E. Craig Dukes}
\author[NIUaddress]{Alexsander Dychkant}
\author[UVAaddress]{Ralf Ehrlich}
\author[NIUaddress]{Kurt Francis}
\author[USAaddress]{M.J. Frank}  
\author[DUBNAaddress]{Vladimir Glagolev}
\author[UVAaddress]{Craig Group}
\author[FNALaddress]{Sten Hansen}
\author[ANLaddress]{Stephen Magill}
\author[UVAaddress]{Yuri Oksuzian}
\author[FNALaddress]{Anna Pla-Dalmau}
\author[FNALaddress]{Paul Rubinov}
\author[DUBNAaddress]{Aleksandr Simonenko}
\author[UVAaddress]{Enhao Song}
\author[UVAaddress]{Steven Stetzler}
\author[UVAaddress]{Yongyi Wu}  
\author[NIUaddress]{Sergey Uzunyan}
\author[NIUaddress]{Vishnu Zutshi}

\address[ANLaddress]{Argonne National Laboratory, Argonne, Illinois 60439, USA}
\address[FNALaddress]{Fermi National Accelerator Laboratory, Batavia, Illinois 60510, USA}
\address[DUBNAaddress]{Joint Institute for Nuclear Research, Dubna, 141980, Russian Federation}
\address[NIUaddress]{Northern Illinois University, DeKalb, Illinois 60115, USA}
\address[USAaddress]{University of South Alabama, Mobile, Alabama 36688, USA}
\address[UVAaddress]{University of Virginia, Charlottesville, Virginia 22904, USA}
\address[TSUHEPIaddress]{High Energy Physics Scientific-Research Institute of Iv. Javakhishvili Tbilisi State University (HEPI-TSU), Tbilisi, 0186, Georgia}

\begin{abstract}

Photoelectron yields of extruded scintillation counters with titanium dioxide coating and embedded wavelength shifting fibers read out by silicon photomultipliers have been measured at the Fermilab Test Beam  Facility using 120\,GeV protons.  The yields were measured as a function of transverse, longitudinal, and angular positions for a variety of scintillator compositions, reflective coating mixtures, and fiber diameters. Timing performance was also studied. These studies were carried out by the Cosmic Ray Veto Group of the Mu2e collaboration as part of their R\&D program.

\end{abstract}


\end{frontmatter}
\today


\section{Introduction}

The Mu2e experiment will search for the neutrino-less conversion of a muon into an electron in the presence of an aluminum nucleus at a single-event sensitivity of about $3{\times}10^{-17}$~\cite{ref:Mu2eTDR}. This represents a sensitivity improvement of four orders of magnitude relative to the current best limit on this process~\cite{ref:SUNDRUM}.  The observation of this process would signal the existence of charged lepton-flavor violation at a level far beyond what is expected from the standard model predictions~\cite{Bernstein:2013hba}.  

A major background for this experiment will be due to cosmic-ray muons that can produce several processes mimicking the signal.  These cosmic-ray induced background events, which will occur at a rate of about one per day, must be suppressed by four orders of magnitude in order to achieve the sensitivity goals of Mu2e.  To do this, an active veto will surround the primary Mu2e detection apparatus on five sides in order to detect penetrating cosmic-ray muons.  The veto will consist of more than 5000 scintillation counters arranged in four layers, each counter is 20\,mm thick by 50\,mm wide and with varying lengths.  This paper describes measurements made with the prototype counters for the Mu2e Cosmic Ray Veto (CRV).

\section{Counter Description}

\begin{figure}[hbt]
\centering
\includegraphics[width=3in]{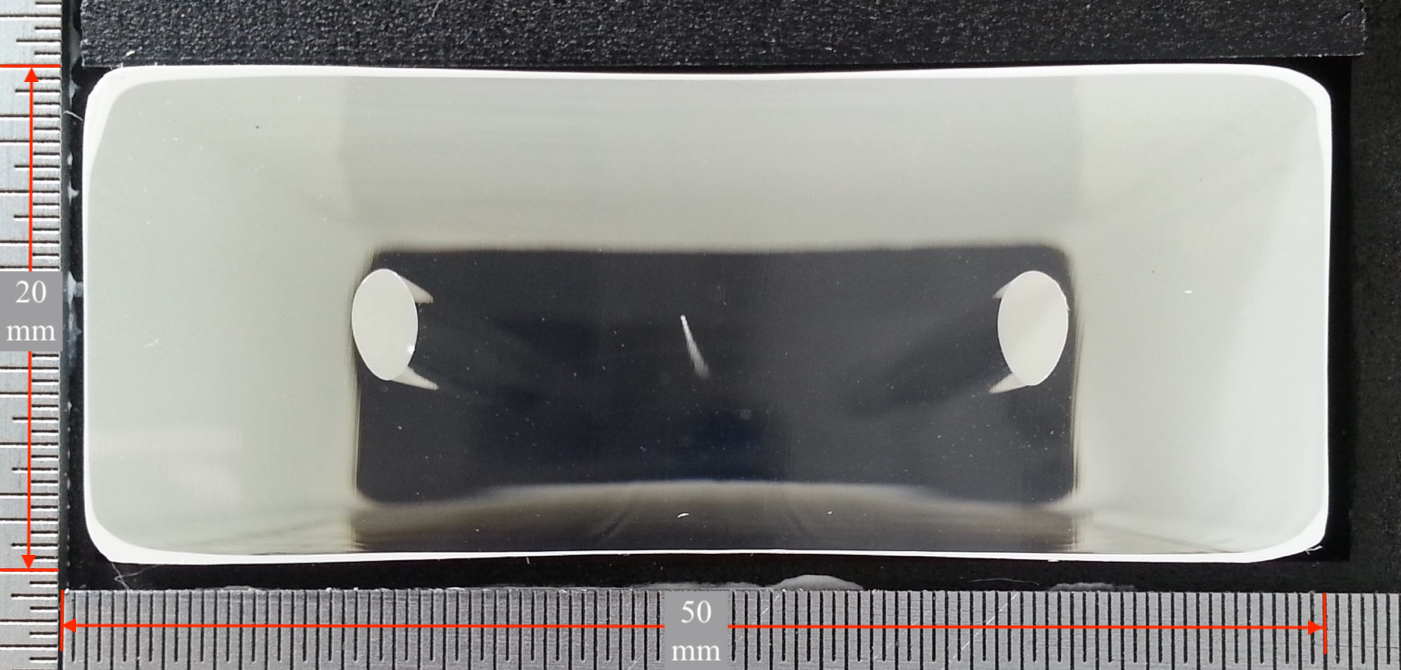}
\caption{Photo from the end of a typical counter.  The shape of the counter, the holes, and the TiO$_2$ coating are visible. Tick marks are spaced by 0.5\,mm.}
\label{fig:counter}
\end{figure}

The counters tested here were all 3000${\times}$50${\times}$20\,mm$^3$. They were extruded at the FNAL-NICADD Extrusion Line Facility \cite{ref:nicadd}. The polystyrene base of each counter was STYRON 665~W. Four different scintillator composition/coating mixtures were tested, and are listed in Table~\ref{table:scint}. The primary dopant was always 2,5-diphenyloxazole (PPO, 1\% by weight). The secondary dopant was either 1,4-bis (5-phenyloxazol-2-yl) benzene (POPOP) or 1,4-bis (2-methylstyryl) benzene (bis-MSB).  A co-extruded reflective coating of 0.25\,mm nominal thickness surrounded the core.  This outer reflective coating was added through material injected from a second extrusion machine (co-extruder) that mixed the polystyrene and TiO$_2$ pellets.  Each counter also had two co-extruded holes of nominal 2.6\,mm diameter into which wavelength-shifting (WLS) fibers were placed.  A cross-sectional view of a counter is shown in Fig.~\ref{fig:counter}, where the shape of the counter, the holes, and the TiO$_2$ coating are visible.

\begin{table}[htbp]
\caption{Scintillator dopants and coatings.}
  \label{table:scint}
\centering
\begin{tabular}{cccc}
\hline
\hline
  \multicolumn{1}{c}{Name} 
& \multicolumn{1}{c}{Primary Dopant}
& \multicolumn{1}{c}{Secondary Dopant}
& \multicolumn{1}{c}{Coating} \\
\hline
A  & 1\% PPO  &  0.03\% POPOP   &  15\% TiO$_2$ \\
B  & 1\% PPO  &  0.03\% POPOP   &  30\% TiO$_2$ \\
C  & 1\% PPO  &  0.05\% POPOP   &  30\% TiO$_2$ \\
D  & 1\% PPO  &  0.05\% bis-MSB &  30\% TiO$_2$ \\
\hline
\hline
\end{tabular}
\end{table}

\begin{figure}[hbt]
\centering
\includegraphics[width=3.5in]{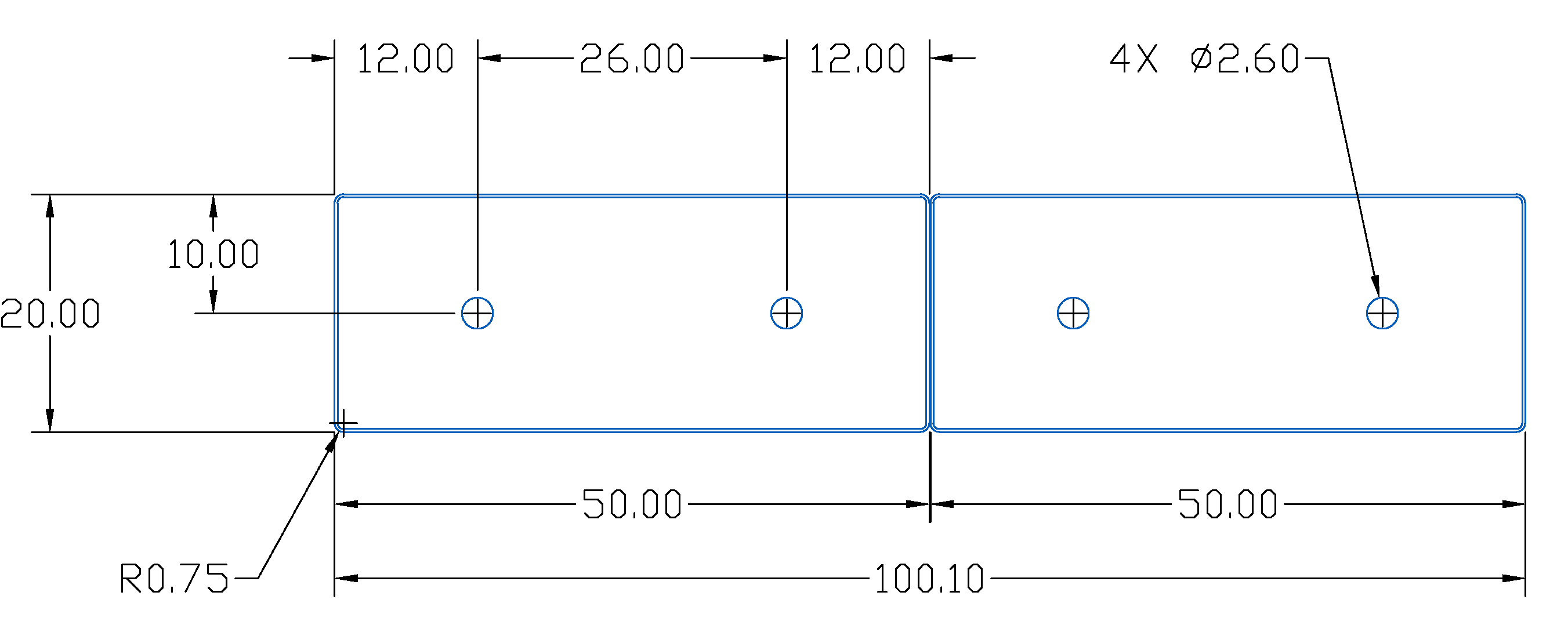}
\caption{Dicounter end view showing fiber positions.  Dimensions (mm) are nominal: actual values are slightly different. }
\label{fig:dicounterend}
\end{figure}

  Counters were assembled into full dicounters at the University of Virginia.  The counters were first glued into pairs called dicounters, using 3M DP420 epoxy \cite{ref:dp420}, producing the profile shown in Fig.~\ref{fig:dicounterend}. The fibers were then placed into the four dicounter holes. The WLS fibers were Kuraray double-clad Y11 doped with 175 ppm K27 dopant, and were non-S-type~\cite{ref:kuraray}. Counters with three different fiber diameters were studied: 1.0, 1.4, and 1.8\,mm.  The fibers were not glued in the extrusion holes, nor were they constrained in any fashion to lie in the holes.   At each end of the dicounter an acetal fiber guide bar was glued to the extrusions using 3M DP100 epoxy~\cite{ref:dp100}. At the same time the WLS fibers were glued into funnel-shaped channels in the fiber guide bars using the same epoxy. The fibers, protruding from both ends of the dicounters, were cut off using a hot knife and the fiber guide bars were then fly cut, which served to polish the fiber ends.

Light captured in the fibers was read out at both ends by $2.0{\times}2.0$\,mm$^2$ (model S13360-2050VE, 1584 pixels, pixel size of 50\,$\mu$) Hamamatsu silicon photomultipliers (SiPMs)~\cite{ref:hamamatsu}. These surface-mount, through-silicon via (TSV) devices were chosen because they have a thin (0.1\,mm) epoxy layer which allows closer proximity between the fiber and photosensor.   Radiation damage from neutrons is a concern.  The devices described in this paper had not been irradiated. Reference~\cite{ref:RadiationNIM} studies the SiPM radiation hardness and its impact on CRV performance.



\begin{figure}[hbt]
\centering
\includegraphics[width=4in]{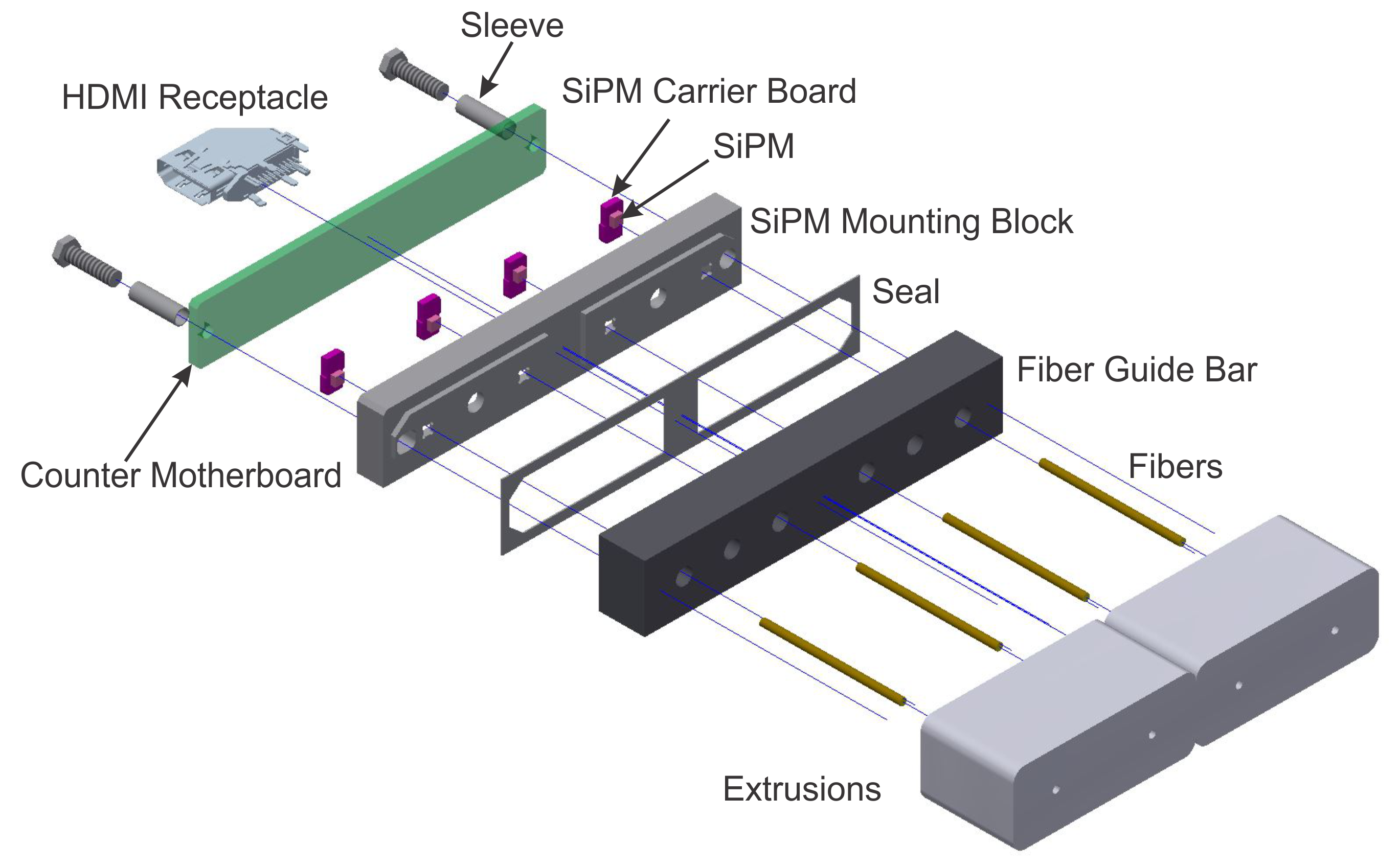}
\caption{Exploded view of the end of a dicounter showing the fiber guide bar, SiPM mounting block, SiPM
carrier boards, SiPMs, and counter motherboard. The flasher LEDs and pogo pins that are soldered to the
counter motherboard are not shown.}
\label{fig:dicounterexploded}
\end{figure}

\begin{figure}[hbt]
	\centering
	\includegraphics[width=0.90\textwidth]{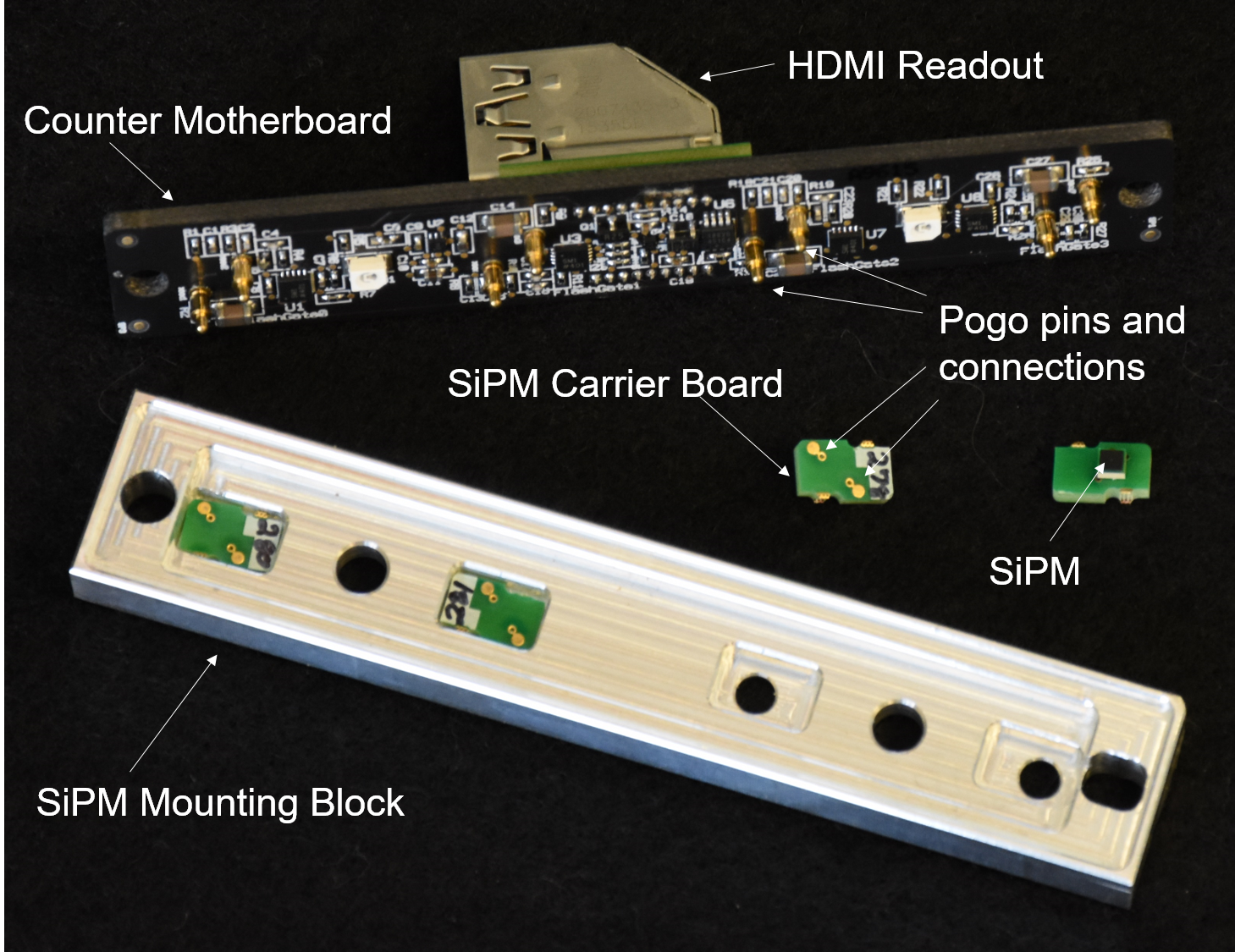}
	\caption{The components used in the electronic readout of the dicounters.}
	\label{fig:manifoldannotated}
\end{figure}

The SiPMs were soldered to small $8.61{\times}5.61$\,mm$^2$ circuit boards, called SiPM carrier boards, that sat in rectangular wells in an anodized aluminum fixture called the SiPM mounting block. Proper registration of the SiPMs to the fibers is critically important in obtaining the maximum light yield, particularly for the 1.8\,mm diameter fibers when mated to the 2.0${\times}$2.0\,mm$^2$ SiPMs. The SiPM mounting blocks were precisely aligned to the fiber guide bars by internally threaded sleeves that were glued into holes on either end of the fiber guide bar. A rubber seal between the fiber guide bar and the SiPM mounting block was used to make the assembly light tight.  An exploded view of the dicounter end is given in Fig.~\ref{fig:dicounterexploded} and a photograph of its components is shown in Fig.~\ref{fig:manifoldannotated}.

The SiPMs were electrically connected via spring-loaded pins (pogo pins) to a small circuit board called
the counter motherboard (CMB). The pogo pins gently pushed down on the SiPM carrier boards, pressing the SiPMs up against the fiber ends. The opaque counter motherboard formed the top of the aluminum SiPM mounting block making the SiPM assembly fixture light tight. The counter motherboard has two flasher LEDs (not used in the tests described here), a thermometer, and an HDMI receptacle. Efforts were made to minimize the extent of the end assembly which, including the HDMI receptacle, still protruded 37\,mm beyond the extrusions.

Signals from the CMB were carried out to a  64-channel front end board (FEB) via a short HDMI cable. The FEB provided bias to the SiPMs, signal pre-amplification and shaping, analog-to-digital conversion at 12.6\,ns intervals (1/79.5\,MHz), and high-speed serial links via Ethernet to a readout controller or a stand-alone computer \cite{ref:feb}.

\section{Experimental Setup}

\begin{figure}[hbt]
\centering
\includegraphics[width=0.99\textwidth]{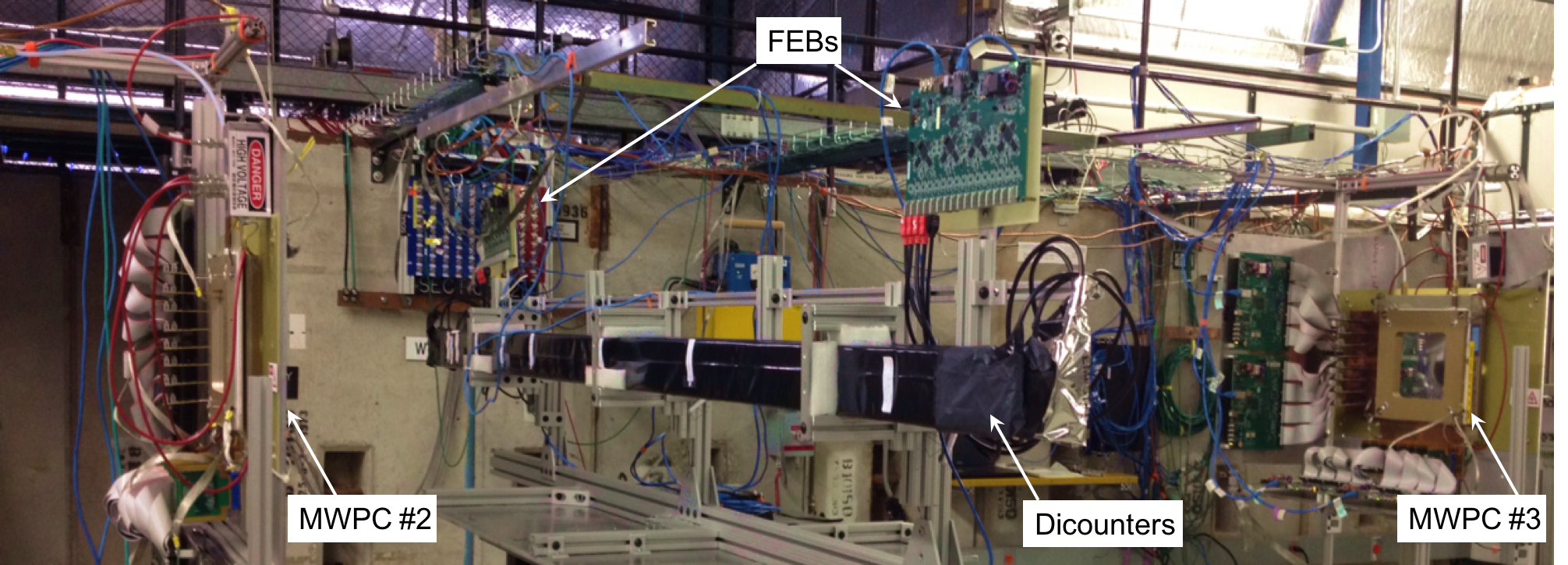}
\caption{Photo of test beam setup.}
\label{fig:TBPic}
\end{figure}

The Fermilab Test Beam Facility is described in Ref.~\cite{ref:ftbf} and a photo of the CRV test-beam setup is shown in Fig.~\ref{fig:TBPic}.  A 120\,GeV proton beam was incident on the counters once a minute in a spill that lasted four seconds.  Up to 3000 events per spill were recorded, so a typical run of 50,000 events took less than 20 minutes.  The counters were mounted on a frame attached to a table with horizontal and vertical motions. Horizontal table motion was only sufficient to position the fixed beam across about half the length of the 3-m-long counters.  Four dicounters were mounted together, one behind the other, so that the beam was incident at the same position on each counter. 

\begin{figure}[h!]
	\centering
        \includegraphics[width=0.85\textwidth]{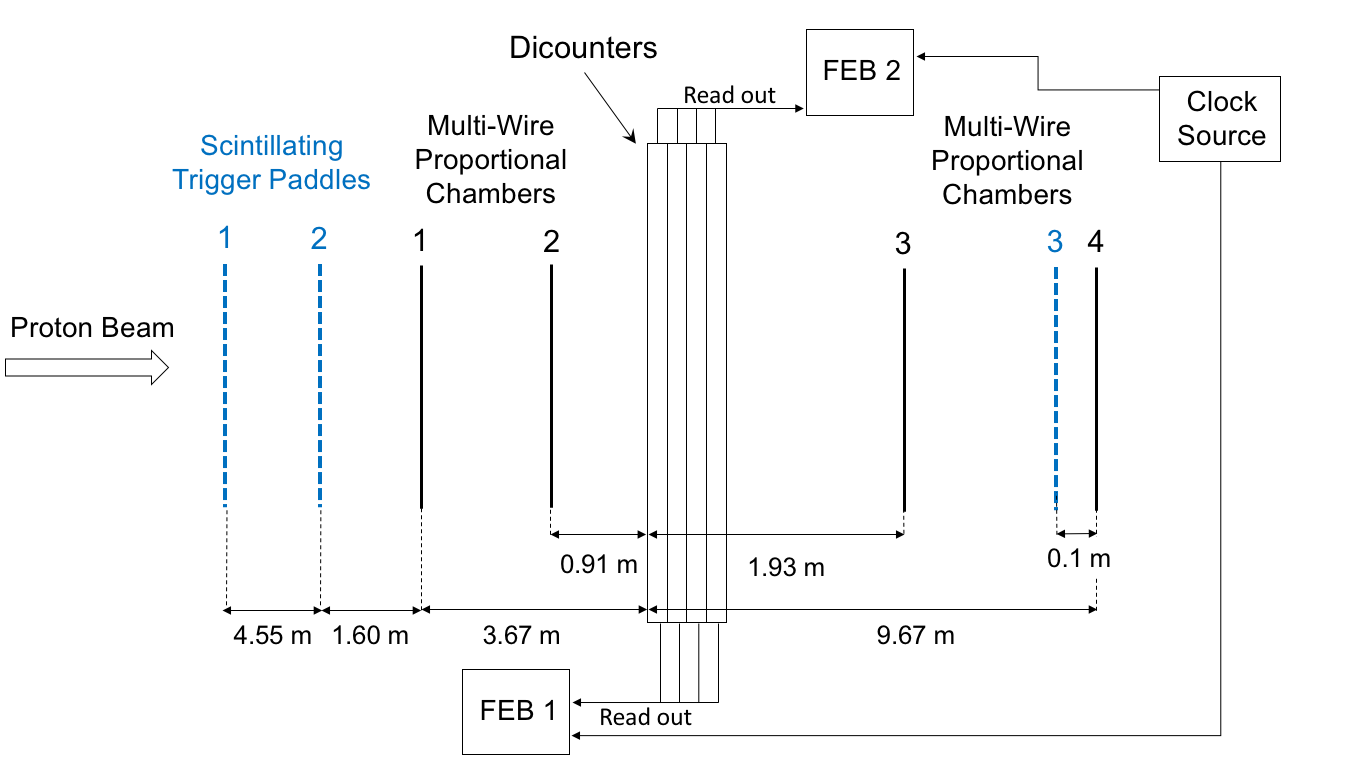}
	\caption{An illustration of the test beam experiment setup (not to scale). In addition to the MWPCs (shown as black lines), the three scintillation counters (shown as blue dashed lines) were also included in the setup for triggering.}
	\label{fig:testbeam_setup}
\end{figure}

\begin{figure}[h!]
	\centering
        \includegraphics[width=0.65\textwidth]{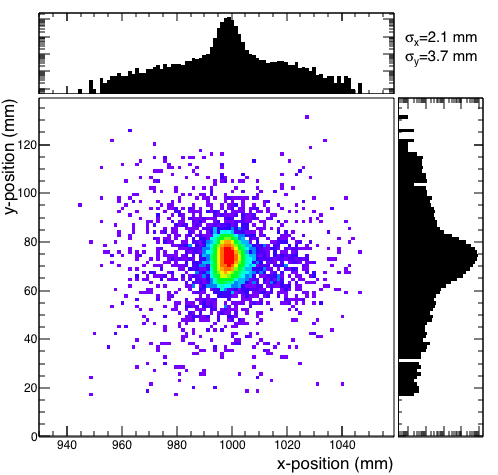}
	\caption{The beam profile for protons which triggered the data acquisition in a typical run. Dicounters in this run were nominally positioned so the proton beam was incident 1000 mm from one end of the dicounter and at 75 mm from the bottom of the dicounter (transversally centered in the top counter).  The x and y projections are shown on a log scale. The Gaussian sigmas from fits to the peak region in the projections are also shown in the figure.}
	\label{fig:beam}
\end{figure}

The experimental setup illustrated in Fig.~\ref{fig:testbeam_setup} employed four 1.0-mm pitch multi-wire proportional chambers (MWPCs) with time-to-digital-conversion readout, two upstream ($z = -3667$\,mm and $z = -906$\,mm) and 
two downstream ($z = 1935$\,mm and $z = 9672$\,mm) of the upstream counters face ($z = 0$\,mm).
Typical residuals were on the order of 0.5\,mm.  Proton tracks were reconstructed from the MWPC data by applying a straight-line fit and requiring a normalized $\chi^2$ value of less than two.  Hits were required in at least three out of the four MWPCs and events with multiple hits in a plane separated by more than four wires were rejected to eliminate events with multiple interactions. The beam profile from a typical run as reconstructed by the MWPCs is given in Fig.~\ref{fig:beam}. The trigger consisted of a coincidence of three scintillation counters, all $101.6{\times}101.6$\,mm$^2$, one upstream, and two downstream of the motion table.

Digitization of the SiPM signals was initiated by the trigger signal from the scintillation counters and a begin-of-spill timing signal from the Fermilab Test Beam Facility.  A total of 127 samples were stored for each triggered event for a total sampling length of 1597\,ns.  The FEB buffered the triggered data during the spill in an on-board DRAM that was read out between spills to a laptop computer.  Two FEBs were used in the data acquisition (DAQ), one for each end of the counters.

Data were taken during two running periods: February and June of 2016. In the February period several fiber diameters were tested including 1.8, 1.4, and 1.0\,mm in different dicounters all with the same scintillator/reflective coating mixture of 1\% PPO + 0.03\% POPOP / 15\% TiO$_2$.  In the June period three different scintillator/reflective coating mixtures were tested, all with 1.4\,mm diameter fibers:
(1) 1\% PPO + 0.03\% POPOP / 30\% TiO$_2$,
(2) 1\% PPO + 0.05\% POPOP / 30\% TiO$_2$, and
(3) 1\% PPO + 0.05\% bis-MSB / 30\% TiO$_2$.
In the analysis described here their light output is always compared to a common reference dicounter from the February 2016 period with 1.4\,mm diameter fibers.

\section{Gain and Photoelectron Calibration}
\label{sec:calibration}

\begin{figure}[h!]
	\centering
	\includegraphics[width=0.6\textwidth]{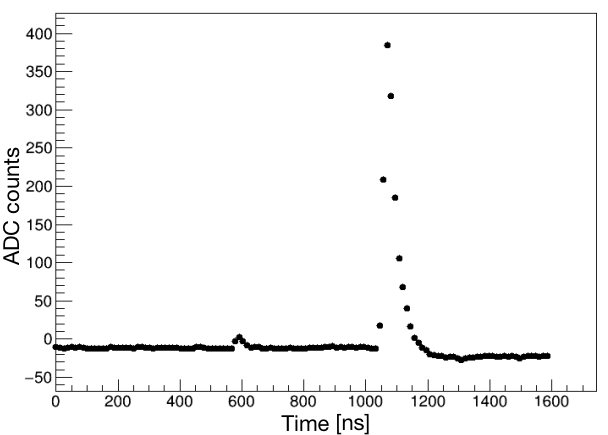}
	\caption{A typical response from a SiPM due to a proton traveling through a counter. The time difference between samples is 12.6 ns. The y-axis is the ADC value digitized by the FEB. Note the single photoelectron noise hit prior to the arrival of the photon and the undershoot after the signal region.}
	\label{fig:adc_vs_tdc}
\end{figure}

SiPMs produce a signal that is proportional to the number of pixels that have fired.  For low-noise and low-crosstalk devices operating with minimal saturation effects, such as the SiPMs used in these measurements~\cite{ref:hamamatsu}\footnote{According to Hamamatsu, the 2.0${\times}$2.0\,mm$^2$ SiPMs operating with a 2\,V overvoltage have less than one million dark noise counts per second above .5 PE, and about 2\% cross talk.}, the number of pixels fired is roughly equal to the photoelectron (PE) yield.   Each SiPM was calibrated to determine how a given response corresponds to the number of photons that were detected.  When a proton traverses a counter, the time-dependent response received from a SiPM will typically look like the signal shown in Fig.~\ref{fig:adc_vs_tdc}. Every 12.6 ns the signal from the SiPM is sampled and digitized.  The resulting ADC value is a function of the gain applied to the SiPM signal by the FEB and the number of SiPM pixels that fired. The response of a SiPM is dependent on the over-voltage, defined as the difference between the applied bias voltage and the breakdown voltage of the SiPM. The breakdown voltage is unique to each SiPM, but it was found that they did not vary by more than $\pm 0.1$\,V. Hence, the same bias was applied to each of the SiPMs. Bias voltages of 55.1 V and 55.3 V were chosen for the February 2016 and June 2016 runs, respectively.  These correspond to an overvoltage of about 2\,V.  The amplifier gain applied at the FEB was set to fill half of the range of the 12-bit ADC.

 A calibration was performed to provide a conversion between an ADC value and the equivalent PE yield.  Before the calibration, the pedestal must be subtracted from all other ADC values in order to zero-center the data. The pedestal was defined to be the most probable ADC value taken from the pre-signal region (0 to 945\,ns or the first 75 samples or in Fig.~\ref{fig:adc_vs_tdc}). Then calibration was done by finding ADC values corresponding to just one or two fired pixels from dark-current pulses.  In Fig.~\ref{fig:adc_vs_tdc}, before the large signal pulse, a small dark pulse is visible corresponding to the signal from a single pixel.

\begin{figure}[h!]
	\centering
        \includegraphics[width=\textwidth]{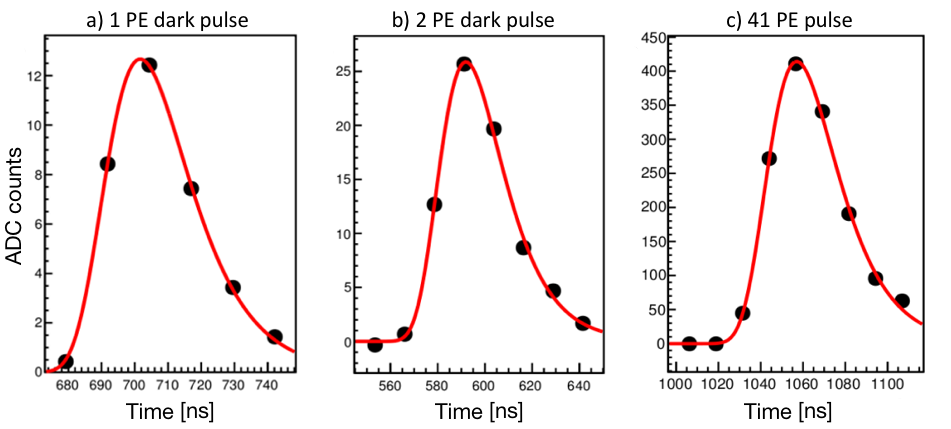}
	\caption{Typical pulses, after pedestal subtraction, and fit with a Gumbel distribution whose integral corresponds to: (a) one photoelectron, (b) two photoelectrons, and  (c) a signal pulse.}
	\label{fig:dark_pe}
\end{figure}

 The integral of each pulse was taken as a measure of the size of the signal.  To do so, a Gumbel distribution~\cite{ref:Gumbel} given by
\begin{equation} \label{gumbel}
	ADC(t) = Ae^{-\frac{t-B}{C} - e^{ -\frac{t-B}{C} }},
\end{equation}
was fit to each pulse\footnote{The Gumbel distribution was selected because it was found empirically to give a good description of the pulse shapes.}. In Eq.~\ref{gumbel}, $A/e$ is the pulse height, $B$ is the pulse time, and $C\pi/\sqrt{6}$ is the pulse width (standard deviation). The integral of Eq.~\ref{gumbel} is given by $A \times C$ and was used as a measure of the total signal output from a SiPM. Figure~\ref{fig:dark_pe} shows fitted pulses corresponding to 1 and 2 photoelectrons due to the dark current.  The figure also includes an example signal pulse.

\begin{figure}[!htb]
		\centering
		\includegraphics[width=.75\textwidth]{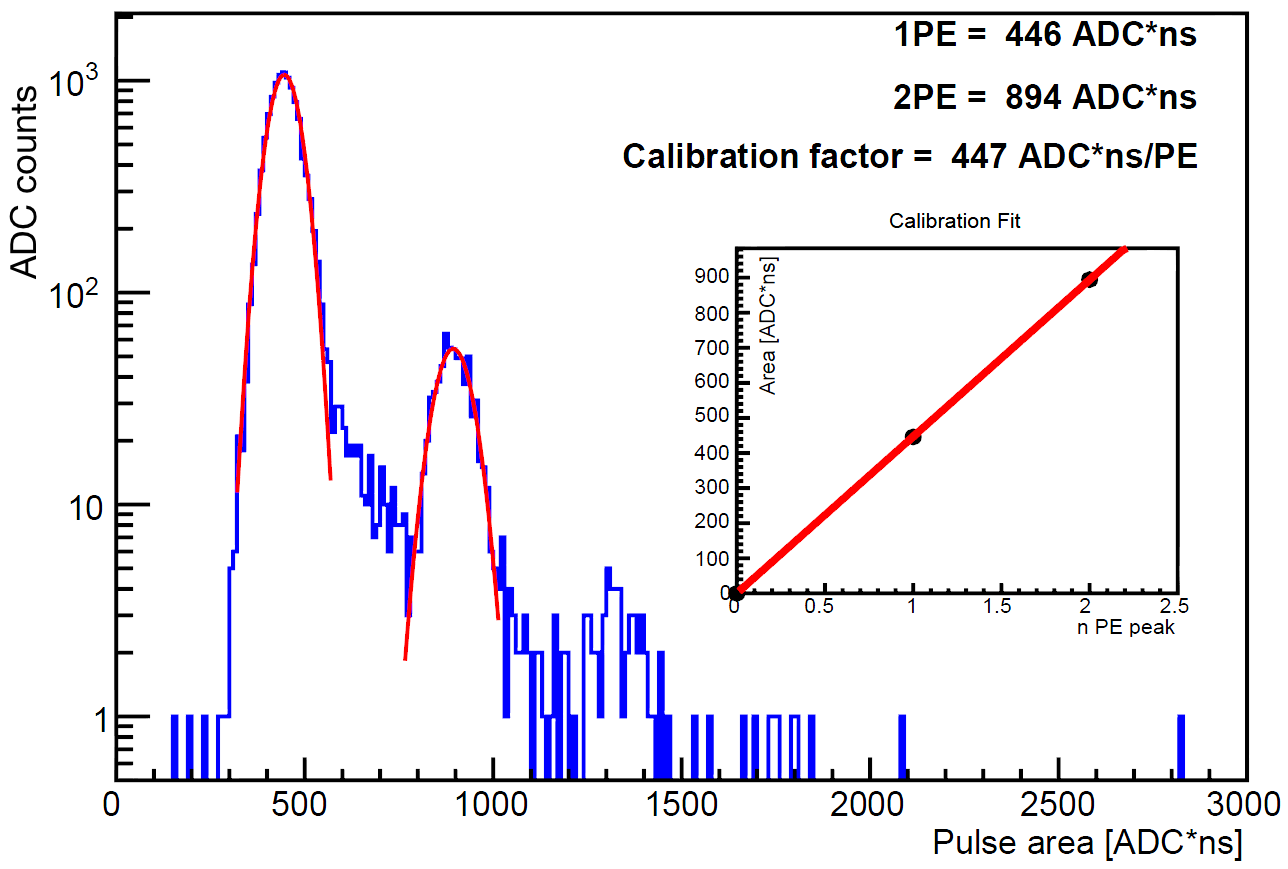}
		\caption{The distribution of the integrals of the pedestal-subtracted dark pulses found in the pre-signal region. Two Gaussian function fits are applied to find the peaks in the histogram which correspond to 1 and 2 PE. The values corresponding to 0, 1, and 2 PE are shown with the black points in the inset. The slope of the fit provides the conversion between signal integrals and PE yield.}
		\label{fig:calibration_hist}
\end{figure}

 To conduct the calibration, the pre-signal region for all events is sampled to find dark pulses like those seen in the pre-signal region of Fig.~\ref{fig:adc_vs_tdc}. All dark pulses found are fit using Eq.~\ref{gumbel} and the distribution of the integrals of these pulses is formed (an example is shown in Fig.~\ref{fig:calibration_hist}). Independent Gaussian fits are applied to the peaks in the histogram and the mean values correspond to the pulse sizes for one and two photoelectrons. A small peak from three photoelectrons can also be observed in Fig.~\ref{fig:calibration_hist}.  However, it was not included in the analysis because the calibration is done in an automated way for all channels and runs and there is not always statistics to resolve this third peak.  The pedestal and the one and two PE values are plotted in the inset of Fig.~\ref{fig:calibration_hist} and a linear fit is applied to the data. The slope of this line provides the conversion between ADC values and PE values for a single SiPM. With this calibration method, conversion factors are stable to better than $\pm 1\%$ between runs under stable temperature conditions ($\pm 0.5^{\circ}$C).

\section{Data Analysis}
\label{sec:analysis}
The test beam data is used to study the PE yield and timing properties of the 3\,m-long counters. Unless otherwise noted, the light yield is determined by summing the PE yield from the two SiPMs on one end of a counter.  The light yield of a counter is determined from the PE yield distribution from each proton that caused a trigger and passed the MWPC event selection. Each PE distribution was fit with the sum of a Gaussian and a Landau function in order to extract the peak and full width at half maximum (FWHM). The peak value of the resulting fit is taken as the most probable PE value and is often referred to simply as the PE yield. 

With $\sim$50,000 events per run the typical statistical uncertainties on the most probable PE values are small ($< 0.1 \%$).   However, in cases where the same counter was measured multiple times from the same beam position, it was observed that PE yield results are stable to $\pm 2$\% or better. Observed variations can be attributed to uncertainties in the calibration, temperature variations, and statistical limitations of the analysis method.\footnote{ The average room temperatures during the February and June data taking periods were 23C and 24C, respectively, with a standard deviation of $<1^{\circ}$C. We did not attempt to correct for temperature dependence, however the analyses that compared PE yield directly between runs were checked, and temperature variations in those runs are stable to $\pm 1^{\circ}$C.  A temperature fluctuation of $1^{\circ}$C corresponds to a photon detection efficiency fluctuation of about 1-2\%.}

\subsection{Studies of Photoelectron Yields}

\begin{figure}[hbt] 
\centering 
\includegraphics[width=0.85\textwidth]{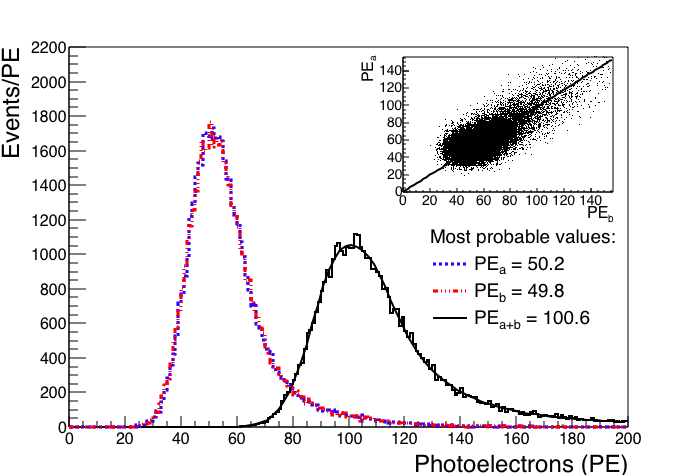}
\caption{Response in photoelectrons at a position 1 m from the SiPM readout for the reference counter read out with 2${\times}$2\,mm$^2$ SiPMs. Dashed and dotted curves are the respective responses from each of the two SiPMs at one end of the counter, and the solid line is the sum of the responses.  The fit described in the text (sum of a Gaussian and a Landau function) is also shown on the summed response.  The inset shows the correlation between the two channels and the line from the correlation fit described in the text.  }
\label{fig:PEYield}
\end{figure}

  Figure~\ref{fig:PEYield} shows the PE distribution of the reference counter when protons are normally incident at the transverse center of the counter and at a position $1$\,m from the SiPM readout.  The figure shows the response for each channel of a counter and the summed response.  An example of the fit described above to extract the most probable PE and the FWHM of the distribution is also shown for the summed response.   Ideally, the two SiPM channels will respond the same way when the proton beam is centered transversally.  The correlation between the two channels is shown in the inset of Fig.~\ref{fig:PEYield}.  A correlation fit of PE$_b$ = $\alpha$ PE$_a$ is applied to the data to find the correlation between SiPM$_a$ and SiPM$_b$. The observed correlation coefficient is $\alpha = 0.9516\pm0.0010$, indicating that the adjacent SiPMs have, on average, closely correlated PE yields. However, there are large statistical fluctuations between the PE yields of the adjacent SiPMs.

  The light output at 1\,m was measured to be about $50~\mathrm{PE}$ per SiPM channel giving the mean of the sum of the two SiPMs of about  $100~\mathrm{PE}$.   These measurements were obtained with the reference counter (type A: scintillator/reflective coating mixtures of 1\% PPO + 0.03\% POPOP / 15\% TiO$_2$), 1.4\,mm diameter fibers, and 2${\times}$2\,mm$^2$ SiPMs.

\subsubsection{Light yield for different scintillator mixes and reflective coating}
\label{sec:scint_compare}

\begin{figure}[h!]
	\begin{minipage}[t]{.52\textwidth}
		\centering 
                \includegraphics[width=\textwidth]{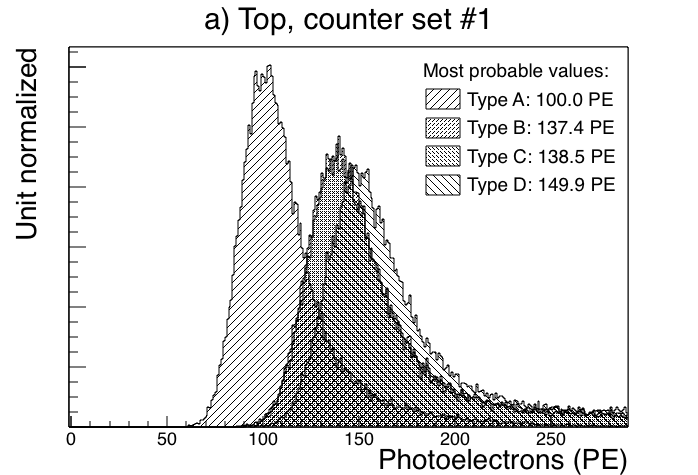}
	\end{minipage}%
	\hspace*{0.01cm}
	\begin{minipage}[t]{.52\textwidth}
		\centering 
                \includegraphics[width=\textwidth]{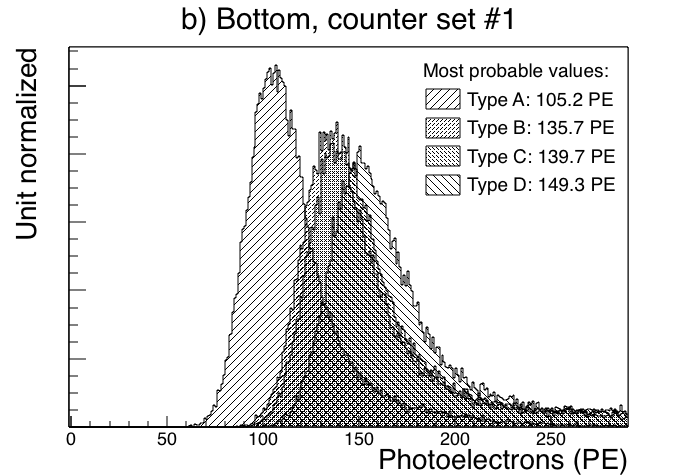}
	\end{minipage}
	\caption{The PE yield distribution for the first counter set shown for the top (a) and bottom (b) counter for each type given in Table~\ref{table:scint}. The proton beam was incident 1\,m from the readout end and centered transversally in the counter. Note that type B and type C are hard to differentiate in the figure due to the similarity of their distributions.}
	\label{fig:counter_config}
\end{figure}

 Counters of the types listed in Table~\ref{table:scint} were placed back to back in the proton beam, as shown in Fig.~\ref{fig:testbeam_setup}. Each counter had SiPMs with size 2$\times$2 mm$^2$ and WLS fibers of 1.4 mm diameter.  A single counter of Type A, which was also used in the February 2016 test-beam run, acted as the reference counter and was the most upstream in each of the June runs. Type B, C, and D counters were placed 2nd, 3rd, and 4th, respectively, in the path of the proton beam. To compare the performance of each counter type, the counters were positioned so the proton beam was normally incident 1\,m from a readout end and transversally centered in either the bottom or the top counter of each dicounter.   Three dicounters of each type were fabricated and the response of each was measured.  The distribution of PE yields for each counter type, is shown in Fig.~\ref{fig:counter_config} for counter set \#1. Table~\ref{table:counter_config} summarizes results of PE yields for each of the counter types for counter set \#1.  The measurements were carried out 1\,m from the other end of the counters as well.  Note that the results from the three counter sets, top and bottom counters, and from the different counter ends all yielded consistent results at the level of better than $\pm$~5\%\footnote{Counter response studies at the University of Virginia CRV factory of hundreds of prototype counters (all of type C) measured a one-sigma variation in counter response of $\sim \pm$~5\%.}.

\begin{table}[h!]
	\centering 
	\caption{PE yield as described in Section~\ref{sec:analysis} for each counter type with 1.4 mm fibers and read out with 2$\times$2 mm$^2$ SiPMs. }
	\begin{tabular}{ccccc}
		\hline \hline
\multicolumn{2}{}{} Counter Set 1 &  & & \\
\hline
	Top & Type & PE Yield & Width & Width / PE Yield \\
& A & 100.0 & 35.0 & 0.35 \\
& B & 137.4 & 42.8 & 0.31 \\
& C & 138.5 & 43.2 & 0.31 \\
& D & 149.9 & 44.6 & 0.30 \\
\hline
	Bottom & Type & PE Yield & Width & Width / PE Yield \\
& A & 105.2 & 37.0 & 0.35 \\
& B & 135.7 & 42.6 & 0.31 \\
& C & 139.7 & 43.2 & 0.31 \\
& D & 149.3 & 45.2 & 0.30 \\
%
%
%

		\hline \hline
	\end{tabular}
	\label{table:counter_config}
\end{table}

 Counter type A, the reference counter, had the lowest response and counter type D had the highest response of the counter types tested. The results from both counter ends were averaged over each of the counter sets and over the data taken when the beam is centered at the top and bottom counter to yield most probable PE values of 101.5, 132.9, 138.0, and 145.6 for types A, B, C, and D, respectively.  Types B, C, and D counters had an average percent light yield increase over the reference counter of $31\%$, $36\%$, and $43\%$, respectively. Types B, C, and D counters also had a smaller width to PE yield ratio when compared to counter type A. However, there was no significant difference in the width to PE yield ratio among counter types B, C, and D.   An important finding is that increasing the amount of TiO$_2$ from 15\% to 30\% in the co-extruded coating resulted in an increase of over 30\% in the PE yield. Changes to the secondary dopant showed an improvement in light yield smaller than 10\%.

\subsubsection{Comparison of light yield for counters with different WLS fiber diameters}

\begin{figure}[h!]
	\begin{minipage}[t]{.52\textwidth}
	    \centering
            \includegraphics[width=\textwidth]{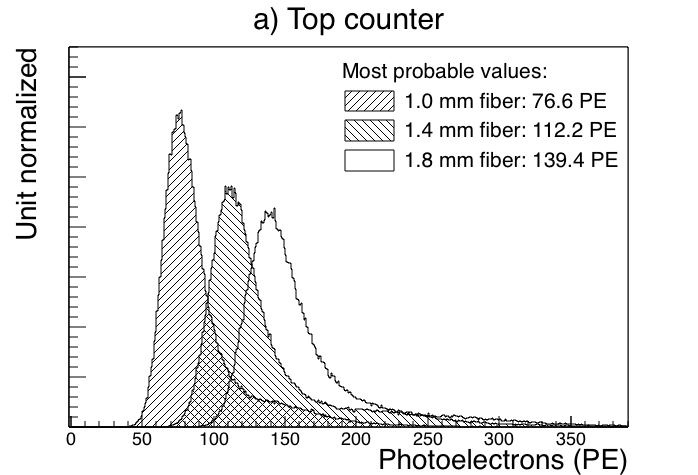}
	\end{minipage}%
	\hspace*{0.01cm}
	\begin{minipage}[t]{.52\textwidth}
	    \centering
            \includegraphics[width=\textwidth]{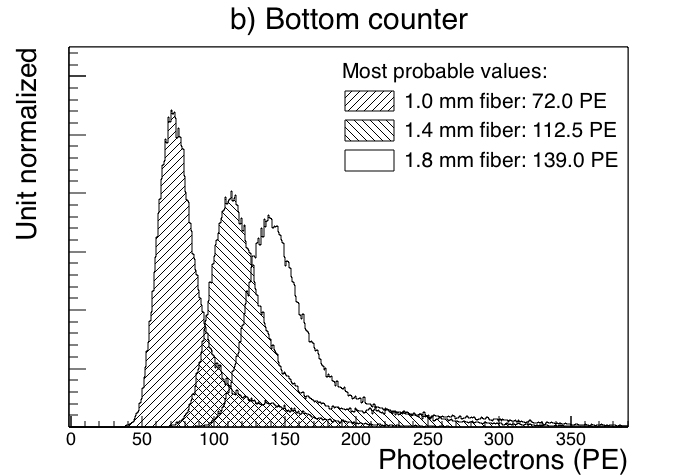}
	\end{minipage}
	\caption{The PE distribution for dicounters with 1.0, 1.4, and 1.8 mm diameter WLS fibers. The proton beam was incident 1 m from the readout end and centered transversally in the  (a) top counter and the (b) bottom counter.}
	\label{fig:fiber_size_top}
\end{figure}

The light yield as a function of the diameter of WLS fibers was studied in the February test-beam run. Three separate dicounters of type A readout with 2$\times$2 mm$^2$ SiPMs were constructed with 1.0, 1.4, and 1.8 mm diameter fibers. In Fig.~\ref{fig:fiber_size_top}, the PE distribution for each of the dicounters is shown for a beam incident 1\,m from the readout end and centered transversally in either the top or bottom counter. A summary of the PE yield for each fiber size is found in Table~\ref{table:fiber_size}.  If the capture of the scintillation light by the fibers is purely a surface effect, it is expected that the PE yield of each fiber size is proportional to the fiber diameter (assuming that the light transmission from the fiber to the SiPM is
independent of the fiber radius).  Similarly, if the light captured by the fibers is purely a volume effect, the light yield would be proportional to the square of the fiber diameter used.

\begin{table}[h!]
\centering
\caption{PE yield as described in Section~\ref{sec:analysis} and fiber size comparison for type A counters read out with 2$\times$2 mm$^2$ SiPMs.  The ratio shown is of the PE yields.}
\begin{tabular}{cccccc}
	\hline
	\hline
  \multicolumn{6}{c}{Top Counter}\\ 
Fiber Size & PE Yield & Width & Width / PE & Ratio$_{1.0}$ & Ratio$_{1.4}$ \\
\hline
1.0 & $ 76.6 $ & $ 28.4 $ & 0.37 & & \\
1.4 & $ 112.2 $ & $ 37.6 $ & 0.33 & 1.47 & \\
1.8 & $ 139.4 $ & $ 44.6 $ & 0.32 & 1.82 & 1.24 \\
\hline
  \multicolumn{6}{c}{Bottom Counter}\\ 
 Fiber Size & PE Yield & Width & Width / PE & Ratio$_{1.0}$ & Ratio$_{1.4}$ \\
\hline
1.0 & $ 72.0 $ & $ 28.0 $ & 0.39 & & \\
1.4 & $ 112.5 $ & $ 37.4 $ & 0.33 & 1.56 & \\
1.8 & $ 139.0 $ & $ 43.6 $ & 0.31 & 1.93 & 1.24 \\
\hline
\hline
\end{tabular}
\label{table:fiber_size}
\end{table}

  The data given in Table~\ref{table:fiber_size} show that PE yield increases roughly linearly with fiber diameter, indicating light yield from the fibers is largely a surface effect.  However, in a study using a 3$\times$3 mm$^2$ SiPM a potential registration problem with the 1.8\,mm fiber was discovered.   It suggests a light loss of about 18\% when the 1.8\,mm fiber is paired with a 2$\times$2 mm$^2$ SiPMs.  If this is taken into account, then the dependence is slightly non-linear, suggesting that light is largely, but not completely, collected near the fiber surface.

\subsubsection{Effect of improving the optical coupling between SiPM and WLS fiber}
\begin{figure}[h!]
    \centering
    \includegraphics[width=0.65\textwidth]{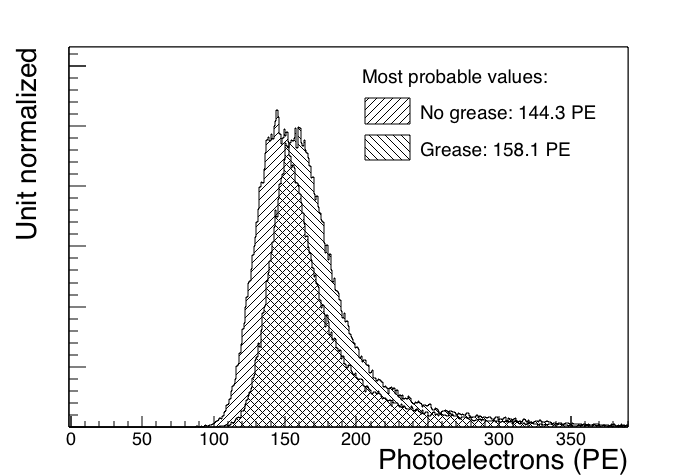}
    \caption{A comparison between the PE distributions for the same counter before and after optical grease was applied between the 1.8 mm diameter fibers and the 2$\times$2 mm$^2$ SiPMs.}
    \label{fig:optical_grease}
\end{figure}

The effect of placing BC-630 Silicone Optical Grease~\cite{ref:grease} (index of refraction of 1.465) between the ends of the WLS fibers and the SiPMs was explored in the February test-beam run. Note that the fiber core has an index of 1.59~\cite{ref:kuraray}  and the epoxy coating of the SiPM has an index of 1.55~\cite{ref:hamamatsu}.  A dicounter of type A with 1.8 mm fibers and 2$\times$2 mm$^2$ SiPMs had optical grease placed between the SiPMs and the fiber ends. Figure~\ref{fig:optical_grease} shows the PE yield of this dicounter when the beam was centered at a position 1\,m from the readout end and centered transversely on the top counter before and after the optical grease was applied. A summary of the results is given in Table~\ref{table:optical_grease}. There is a small (10\%) increase in the PE yield due to the application of optical grease indicating that the fiber coupling to the SiPM without the grease is already good.

\begin{table}[h!]
\centering
\caption{Summary of change in PE yield as described in Section~\ref{sec:analysis} due to application of optical grease between the 1.8 mm diameter fibers and the 2$\times$2 mm$^2$ SiPMs.  The ratio shown is of the PE yields.}
\begin{tabular}{ccccc}
	\hline
	\hline
	 & PE Yield & Width & Width / PE Yield & Ratio\\
	 \hline
No Grease & $ 144.3 $ & $ 45.4 $ & 0.31 & \\
Grease & $ 158.1 $ & $ 46.2 $ & 0.29 & 1.1\\

	\hline
	\hline
\end{tabular}
\label{table:optical_grease}
\end{table}

\subsubsection{Studies of reflectors on counter ends}
\label{sec:reflector}
Dicounters with electronic readout on both ends produce some level of reflection due to the previously mentioned change in the index of refraction between the fibers, air, and SiPMs. For counters with single-ended readout, the PE yield can be enhanced by adding reflectors to the far end.  Using single-ended readout may be cost effective, or may be required due to issues of access, space constraints, or high radiation levels.

\begin{figure}[h!]
	\begin{minipage}[t]{.52\textwidth}
	    \centering
		\includegraphics[width=\textwidth]{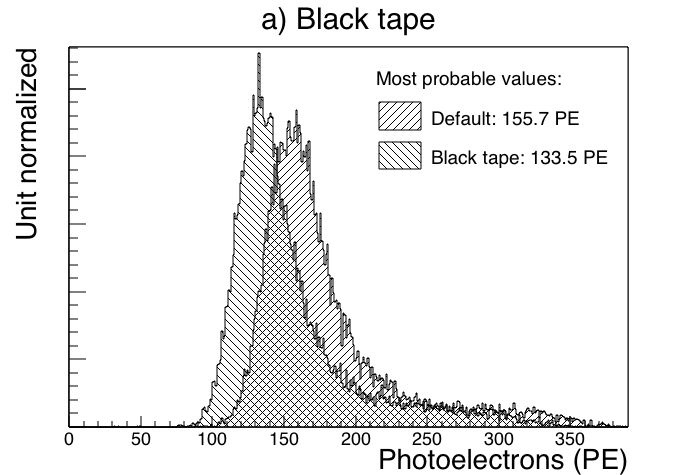}
	\end{minipage}%
	\hspace*{0.01cm}
	\begin{minipage}[t]{.52\textwidth}
	    \centering
		\includegraphics[width=\textwidth]{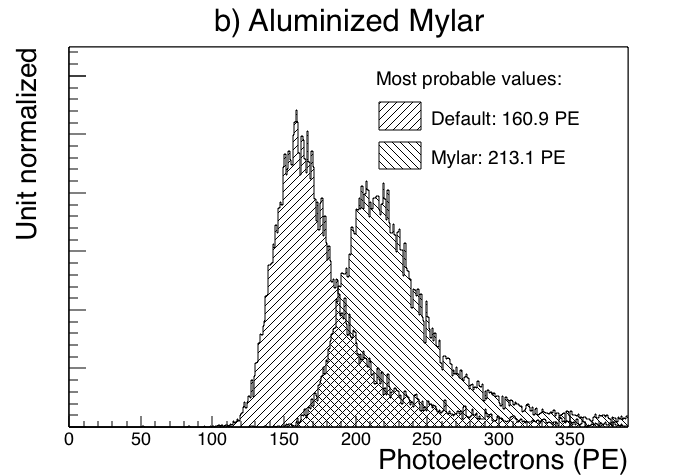}
	\end{minipage}
	\caption{A comparison of the PE distributions obtained with the beam at 1\,m from a readout end from the same counter with default readout and after (a) black tape or (b) aluminum-coated Mylar was applied to the opposite end of the counter.  The default case refers to readout electronics (SiPM) on the far end. The beam was located 1\,m from the readout end of the dicounter. }
	\label{fig:reflectivity}
\end{figure}

In order to measure the effect of reflection at the opposite end of a counter on PE yields, two different dicounters of type A with fiber diameters 1.8\,mm and 2$\times$2 mm$^2$ SiPMs were modified by replacing the electronic readout at one end with a reflective material. Black tape was applied at one end of the counter to set a baseline for the PE yields without reflection. Another dicounter had one end capped with aluminum coated Mylar. In Fig.~\ref{fig:reflectivity}(a) and Fig.~\ref{fig:reflectivity}(b), the PE distributions are plotted when the beam was incident 1\,m from the readout end and centered transversally on the top counter of each dicounter tested. In the figure, one distribution comes from a run of the same dicounter with SiPMs at both ends and one comes from a run with one of the described modifications at one end. A summary of the results is given in Table~\ref{table:reflectivity}.

\begin{table}[h!]
\centering
\caption{PE yield as described in Section~\ref{sec:analysis} with different reflector modifications. The ratio shown is of the PE yields.}
\begin{tabular}{ccccc}
	\hline
	\hline
				& PE Yield & Width & Width / PE & Ratio to default\\
				\hline
Default & $ 155.7 $ & $ 48.4 $ & 0.31 & \\
Black Tape & $ 133.5 $ & $ 41.6 $ & 0.31 &  0.86\\
\hline
Default & $ 160.9 $ & $ 45.6 $ & 0.28 & \\
Mylar & $ 213.1 $ & $ 58.0 $ & 0.27 & 1.32\\
\hline
\hline
\end{tabular}
\label{table:reflectivity}
\end{table}

  From Table~\ref{table:reflectivity}, the black tape reduced the PE yield by 14\% compared to SiPM readout electronics on the far end. The aluminum coated Mylar increased PE yield by 32\% relative to the readout electronics.\footnote{A more thorough study of various reflectors has been conducted using a dark box and radioactive source by a subset of the authors of this work and will be published soon.}

\subsection{Longitudinal, Transverse, and Angular Counter Scans}

\begin{figure}[hbt]
\centering
\includegraphics[width=3in]{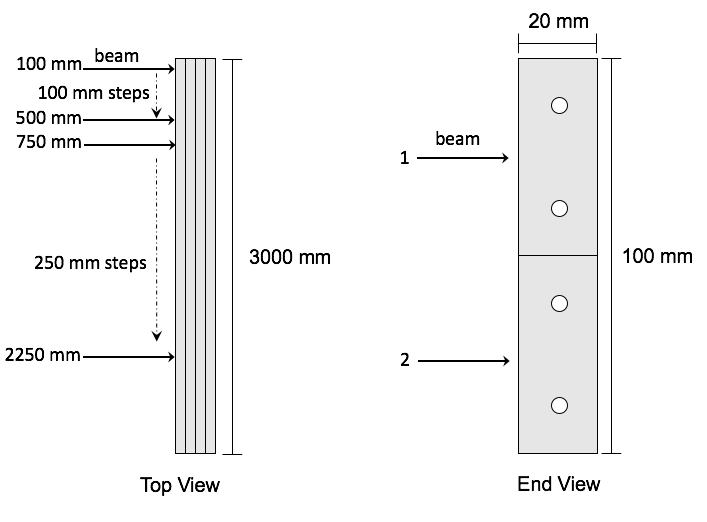}
\caption{Beam longitudinal and transverse positions used in the counter scans.}
\label{fig:BeamScans}
\end{figure}

 Data were taken with the proton beam positioned 100\,mm from one end to 750\,mm from the opposite end of the $3~\mathrm{m}$ long dicounters. These longitudinal scans were taken with the beam transversally centered in each counter in the dicounter.  Near one counter end, several runs were also taken to study the falloff of the response due to counter end effects.    Figure~\ref{fig:BeamScans} illustrates target locations of the proton beam.  The counter response was also studied in four angular orientations relative to the beam.

\subsubsection{Longitudinal scans}
\label{sec:long_scans}
\begin{figure}[h]
    \centering
    \includegraphics[width=0.70\textwidth]{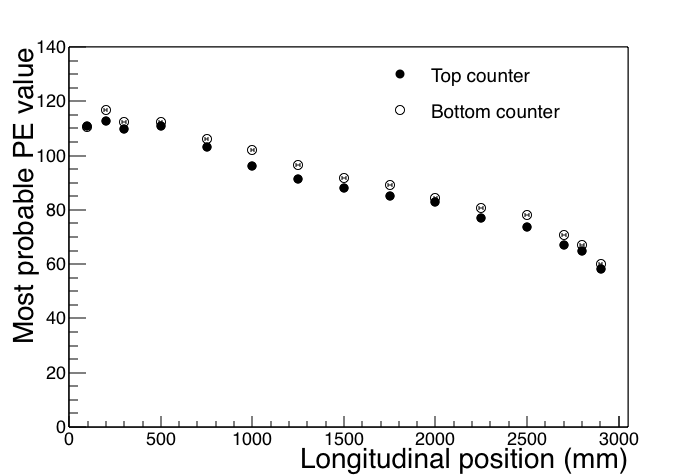}
    \caption{ The PE yield (most probable value) of the standard counter as a function of the distance from the SiPM readout. Solid (open) circles are data from the top (bottom) counter. The statistical uncertainties on the most probable values of the PE distributions are too small to be visible on the figure.}
    \label{fig:attenuation_standard}
\end{figure}

The results reported previously all had the beam positioned at 1\,m from the readout end. Longitudinal scans were made to study the variation in PE yield with distance from the readout end.   For these studies, multiple runs were made with the proton beam normally incident on various positions along the longitudinal length of the dicounter and transversally centered either in the top or bottom counter (see Fig.~\ref{fig:BeamScans}).  The motion table, to which the dicounters were fixed, did not allow for further movement beyond 2250\,mm. However, since data are read out from both ends of the dicounter, data is available for beam positions from 100 to 2900\,mm (distance from one end of the dicounter of 100\,mm implies a distance from the other end of 2900\,mm). Figure~\ref{fig:attenuation_standard} shows the PE yield of the reference counter of type A with 2$\times$2 mm$^2$ SiPMs and 1.4\,mm diameter fibers as a function of the beam position.

The data points for the top and bottom counters in Fig.~\ref{fig:attenuation_standard} follow the same trend. In principle, the data in Fig.~\ref{fig:attenuation_standard} can be fit with an exponential function to estimate the attenuation length in the fiber.  In practice, the result is sensitive to the range used in the fit due to edge effects in the counters (light loss from the end of the counter increases when the proton beam is closer to the counter end).  In addition, the light attenuation in the fiber is wavelength dependent causing the average attenuation length to vary depending on the distance of the incident particle from the readout.  Finally, the attenuation measurement is sensitive to the presence of light reflected from the far end of the fiber which is significant for the SiPM readout as discussed in section~\ref{sec:reflector}.  In Ref.~\cite{ref:fiber} the attenuation length was measured as a function of wavelength in the fiber using an apparatus designed for this purpose.

\subsubsection{Longitudinal scan near counter end}
\label{sec:endscan}
\begin{figure}[h!]
    \centering
    \includegraphics[width=0.7\textwidth]{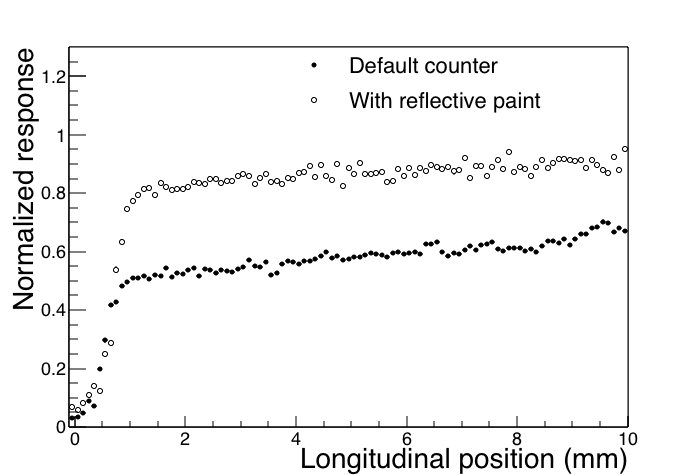}
    \caption{The falloff of the PE yield in the rough-cut counter (default) compared to the counter whose ends were painted with reflective paint.  Counters were normalized to their own response far from the readout end ($80<x<120$ mm) so that the falloff near the end could be directly compared. Statistical error bars are too small to be visible on the figure. }
    \label{fig:EndData}
\end{figure}

  Fine scans near the readout end of counters were conducted in order to study the falloff of the response due to light escaping from the end of the counter.  The default counters were rough cut with a circular saw before the black fiber guide bar was epoxied on.  A special counter was also prepared in which both counter ends were painted with BC-620 reflective paint from Saint-Gobain~\cite{ref:grease} before attaching the fiber guide bar.  The falloff of the PE yield for the counter with reflective paint was compared to the reference counter.  Data from several runs were used with the beam incident between 0 and 100\,mm from the counter end.  Each channel's response was normalized to the value far from the readout end ($80<x<120$ mm).    Consistent behavior in the four SiPM channels (two near and two far) was observed for each counter and the data were combined into the summary plot shown in Fig.~\ref{fig:EndData}.  Both counters show a falloff in the PE yield as the protons get closer to the counter end.  However, the painted counter performs significantly better.  For protons 2\,mm from the counter end the reference counter has lost about 50\% of its light yield, while the counter with painted ends has only lost about 20\%.   Beyond $100$\,mm there is no noticeable difference ($<2$ \%) between the performance of the reference counter and the one whose end was prepared with white reflective paint.

\subsubsection{Transverse response}

The response of the dicounters along their 100\,mm transverse width was measured for the reference counter when the proton beam was incident 1.5\,m from the SiPM readout. Even though the proton beam was incident either 25\,mm (center of bottom counter) or 75\,mm (center of top counter) from the bottom of the dicounter, the spread of the proton beam spot shown in Fig.~\ref{fig:beam} allows for the transverse response to be studied over the entire counter\footnote{Note that a more detailed transverse scan was produced in a past test-beam study and reported in Ref.~\cite{ref:TB2015}.}.  To increase the number of events across the transverse width, the response at both ends of the fiber was combined for six runs: three when the beam was centered on the top counter and three for the bottom counter.

\begin{figure}[h!]
    
    \begin{minipage}[t]{.52\textwidth}
      \centering
    \includegraphics[width=\textwidth]{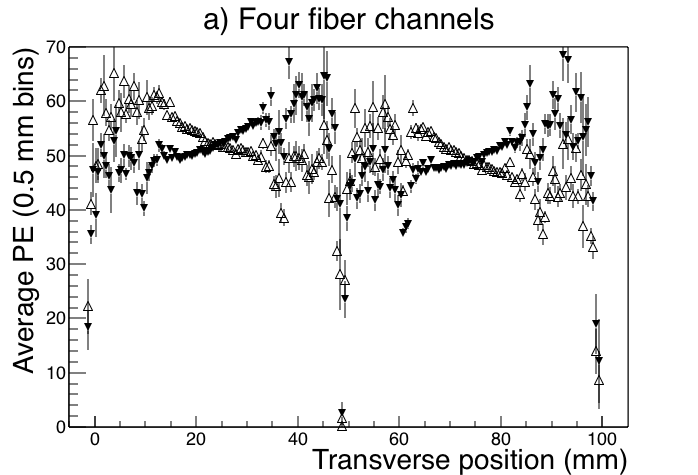}
    \end{minipage}
\hspace*{0.01cm}
    \begin{minipage}[t]{.52\textwidth}
      \centering
    \includegraphics[width=\textwidth]{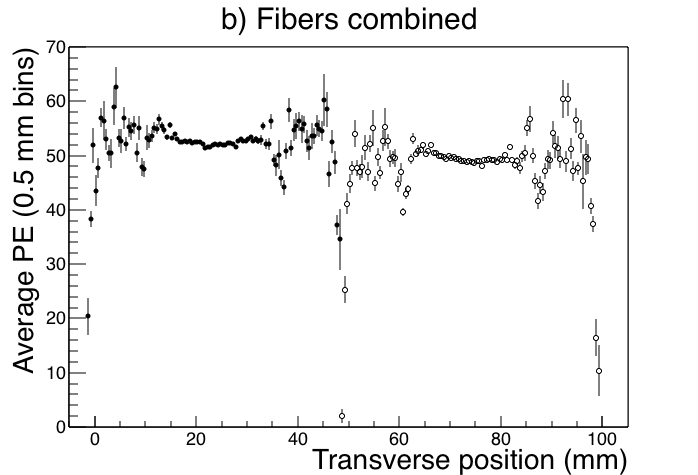}
  \end{minipage}
    \caption{The average PE yield in the default counter type for (a) each of the four SiPM channels on one end of a dicounter and (b) averaged over the fibers in each counter, both shown as a function of the transverse position of the incident proton.  In (a) the open (closed) triangles show the bottom (top) fiber in each counter in a dicounter (bottom counter between 0 and 50\,mm  and top counter between 50 and 100\,mm).  The beam was located 1.5\,m from the end of the dicounter. The error bars are large in locations where there were few incident protons.}
    \label{fig:transverse}
\end{figure}

 The average PE yield as a function of position across the transverse width for each fiber of the dicounter is shown in Fig.~\ref{fig:transverse}(a).  For a given fiber there is approximately a 20\% variation in light yield across the counter.  This is expected due to the relatively poor light attenuation in the scintillator extrusions and the shadowing effect of the fiber channels.

Combining the fibers in each counter, the average PE yield as a function of position across the transverse width of the dicounter is shown in Fig.~\ref{fig:transverse}(b).  
An important feature of the data in Fig.~\ref{fig:transverse} is the sharp drop off in average PE yield when the protons are close to the boundaries of each counter. Since each counter is 50\,mm wide, these boundaries occur at nominal positions of 0, 50, and 100\,mm. There is an effective gap of approximately 0.5\,mm between the two counters due to the TiO$_2$ coating.  There is also a dip in the average PE yield where the WLS fibers are located in each counter, at positions of $\sim$ 10, 35, 60, and 85\,mm.   This is consistent with the protons passing through less scintillating material due to the channels housing the fiber.

\subsubsection{Angular scans}   

\begin{figure}[hbt]
\begin{minipage}[t]{.52\textwidth}
      \centering
\includegraphics[height=0.68\textwidth]{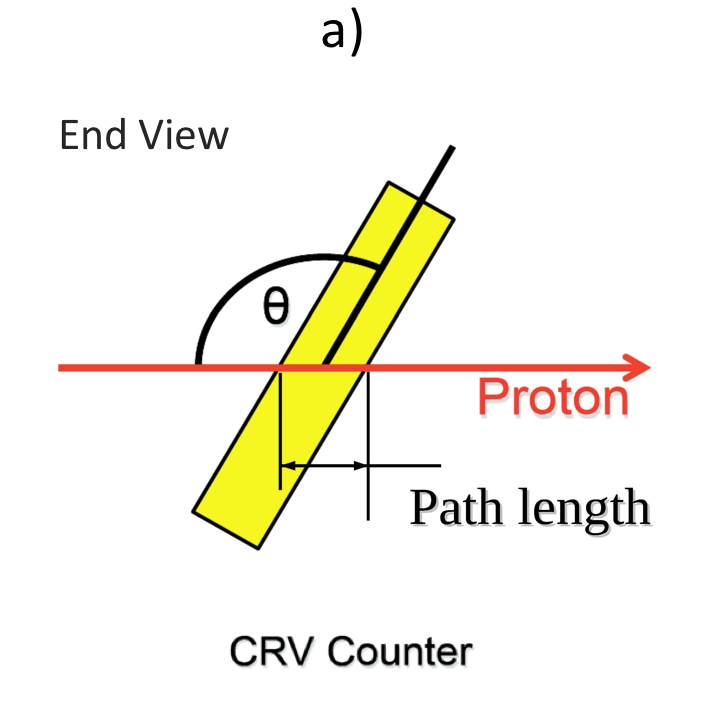}
    \end{minipage}
\hspace*{0.01cm}
    \begin{minipage}[t]{.52\textwidth}
      \centering
\includegraphics[width=\textwidth]{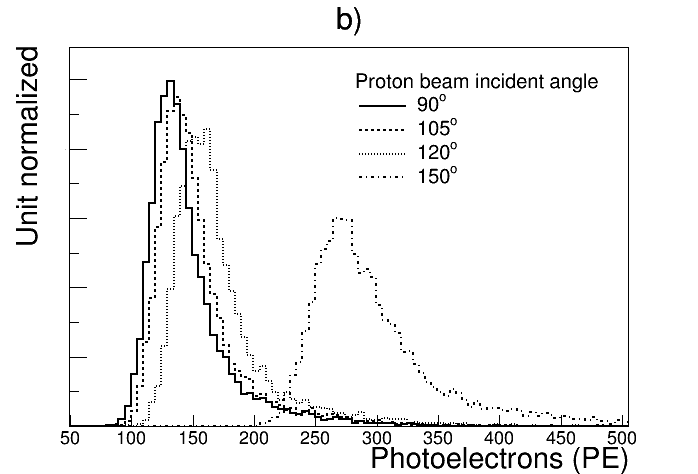}
  \end{minipage}
\caption{ (a) The CRV counter orientation during angle scans with the proton beam 1 m from the counter end. 
(b) PE yields corresponding to the four angular positions at $\theta= 90^\circ $(nominal), 
105$^\circ $, 120$^\circ $, 150$^\circ $ measured for the counter with 1.8\,mm diameter fibers.}
\label{fig:angle-scan1}
\end{figure}

\begin{figure}[hbt]
\begin{minipage}[t]{.52\textwidth}
    \centering
\includegraphics[width=\textwidth]{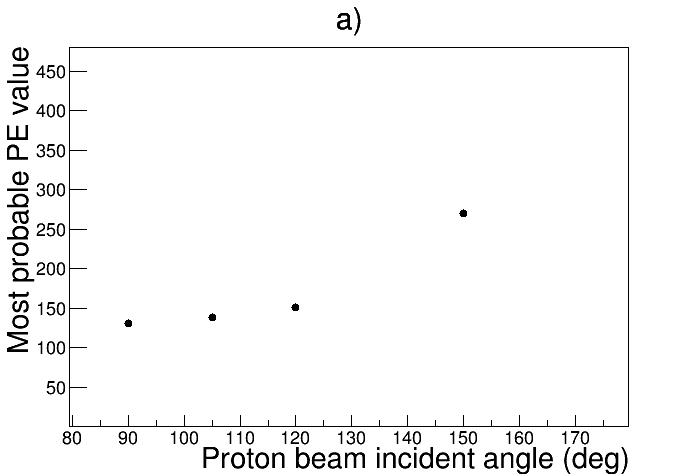}
  \end{minipage}
\hspace*{0.01cm}
  \begin{minipage}[t]{.52\textwidth}
    \centering
\includegraphics[width=\textwidth]{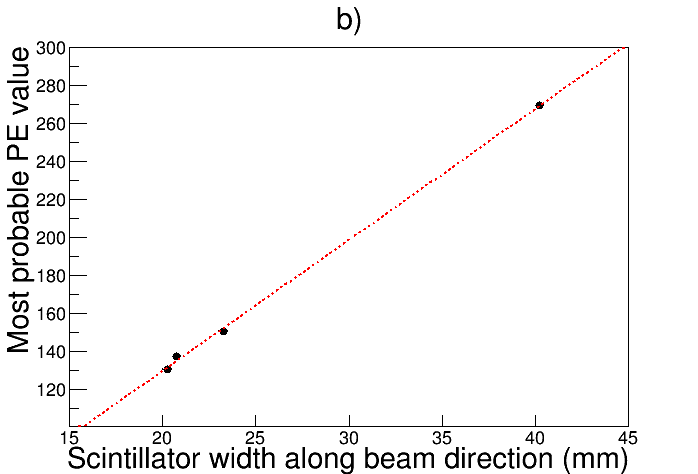}
\end{minipage}
\caption{ The most probable PE value obtained as a function of (a) counter angular position and (b) the proton path length through the counter. }
\label{fig:angle-scan2}
\end{figure}

The counter response was studied at four angular orientations of the incident proton beam relative to the counter $\theta= 90^\circ $, 105$^\circ $, 120$^\circ $, and 150$^\circ $ as illustrated in  Fig.~\ref{fig:angle-scan1}(a). In the longitudinal dimension protons entered a counter  at 1\,m from the edge.  The corresponding PE responses are shown in Fig.~\ref{fig:angle-scan1}(b) for the counter with a 1.8\,mm diameter fiber. The PE yields and widths of the fits for this data are shown in Fig.~\ref{fig:angle-scan2} as functions of the angle $\theta$ and the corresponding width of the counter along the beam direction. A linear increase of  the number of photoelectrons as a function of the incident proton path length through the scintillator is observed.  

\subsection{Timing Studies}

\subsubsection{Single-channel time resolution studies}

\begin{figure}[h!]
	\centering
	\begin{minipage}[t]{.52\textwidth}
		\centering
		\includegraphics[width=\textwidth]{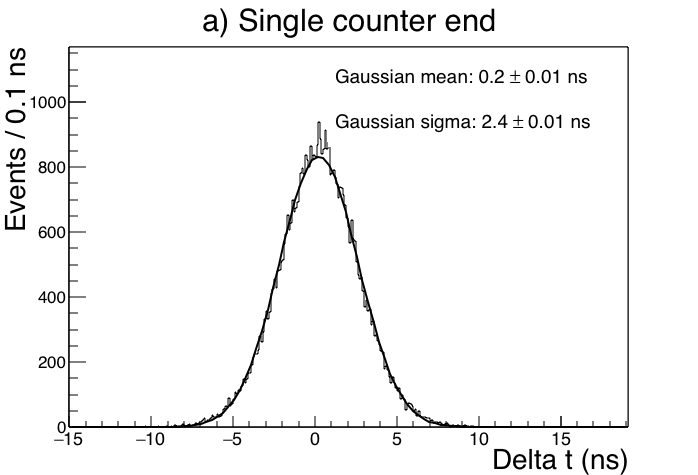}
			
	\end{minipage}%
	\hspace*{0.01cm}
	\begin{minipage}[t]{.52\textwidth}
		\centering
		\includegraphics[width=\textwidth]{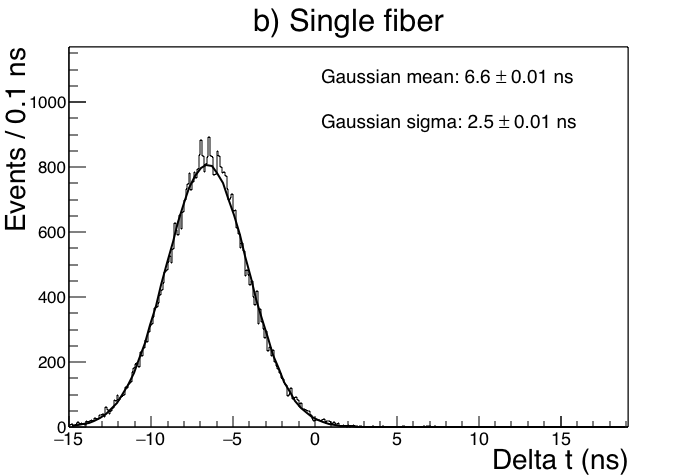}
		
	\end{minipage}
	\caption{The distribution of the time difference measured in the two channels on one end of (a) a single counter and (b) from both ends of a single fiber. }
	\label{fig:TimingResolution}
\end{figure}

The photon arrival time at a SiPM is calculated by using the Gumbel fit described in Sec.~\ref{sec:calibration} to estimate the peak time of the pulse\footnote{The peak time, rather than the leading edge time, will be used by the CRV because the peak time method is less sensitive to the expected rate of noise hits in the counters due to the neutron and gamma backgrounds in the detector hall.}.  The single-channel time resolution can be estimated from the time difference between two channels on the same side of the same extrusion, since the incident proton should produce a coincident pulse in both channels. The standard deviation of the time difference in the channels $\sigma_{\mathrm{diff}}$ is related to the single-channel time resolution $\sigma_{\mathrm{res}}$ by a factor of $\sqrt{2}$ ($\sigma_{\mathrm{res}}=\frac{\sigma_{\mathrm{diff}}}{\sqrt{2}}$).  Figure~\ref{fig:TimingResolution}(a) shows the distribution of the time difference measured between the two channels of one counter end.  The distribution is peaked at zero with a Gaussian sigma of 2.4\,ns indicating a single channel timing resolution of 1.7\,ns.  The mean offset of 0.2\,ns is small compared to the resolution and is probably due to differences in the routing of the two channels in the electronic readout.  The result in Fig.~\ref{fig:TimingResolution} uses counter type C.  For the reference counter, type A, the timing resolution is 1.9\,ns, while for counter type D the resolution is 1.6\,ns.

Figure~\ref{fig:TimingResolution}(b) shows the distribution of the time difference measured in the two channels from opposite ends of the same fiber in a counter of type C when the beam is located at the longitudinal center of the counter.  In this case, the fiber ends were read out with two different FEBs.  The resolution is close to that observed in Fig.~\ref{fig:TimingResolution}(a) where a single FEB was used.  The results suggest that routing in the electronic readout does not have a large impact on the time resolution and that time resolution is primarily limited by variations of photon path lengths in the scintillator and fiber affecting the mean arrival time at the SiPM.  The large offset in the mean of Fig.~\ref{fig:TimingResolution}(b) will be discussed in the next section.

\subsubsection{Speed of light in the fiber}
The difference in the arrival times of photons at the SiPMs on opposite ends of the fibers can be used to determine the speed that the light signal travels down the WLS fibers.   The two ends of a dicounter are connected to two separate readout boards (FEBs) that are synchronized to a 26~MHz reference clock fanned out from a controller module. There are timing offsets between the cards that arise from a variety of sources including differences in cabling lengths and logic delays. See Fig.~\ref{fig:TimingResolution}(b) for an example of this offset before a correction is applied.  In order to find the time offsets and make a correction, the average time difference between left and right readouts was found when the beam was positioned in the middle of the length of the dicounter, 1500\,mm from either end.  The average time difference varies slightly by channel, but was found to be about 6\,ns.

  The path difference for light traveling to both ends of a fiber from a particle incident at position $x$ is $L-2x$, where $L$ is the length of the counter and $x=0$ is defined to be the end of the counter.  So, if the light arrival time difference $\Delta t$ is detected between SiPMs at opposite ends of the counter, the incident position of the proton along the counter length can be calculated as
\begin{equation} \label{eqn:x_position}
    x = \frac{L - \Delta t * v_{\gamma}}{2}.
\end{equation}
Hence, the speed of light in the fibers, $v_{\gamma}$, is

\begin{equation}
    v_{\gamma} = \frac{L - 2x}{\Delta t}.
\end{equation}

\begin{figure}[h!]
    \centering
    \includegraphics[width=0.65\textwidth]{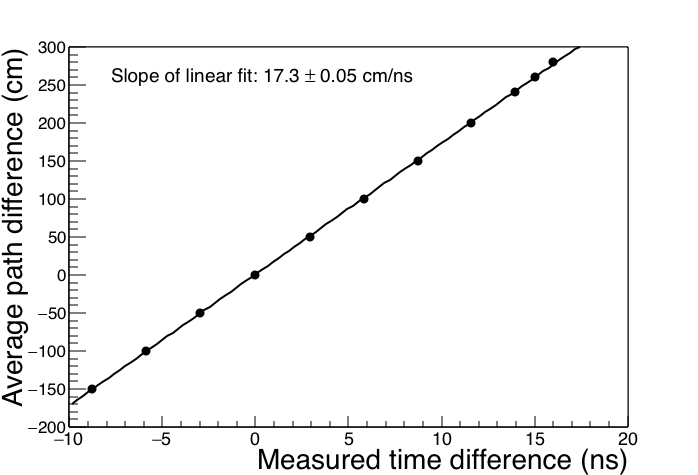}
    \caption{The average path length difference as a function of the average corrected arrival time difference of photons from proton traversal signals collected from both ends of a single fiber.  Statistical errors on the time difference are taken as the error on the mean from the Gaussian fit and are too small to be visible on the figure.  }
    \label{fig:speed_of_light}
\end{figure}

Figure~\ref{fig:speed_of_light} shows an example from one fiber of the average path length difference based on the MWPC position versus the average corrected arrival time difference of photons arriving at each end of the fiber. Data was gathered when the beam was transversally centered on the top counter and at various positions along the longitudinal position of the counter, allowing for multiple path length differences to be sampled.  Applying a linear fit to the data gives a speed of light in the WLS fibers of $17.3$ cm/ns ($0.58c$), where $c$ is the speed of light in vacuum.   Using Kuraray's quoted index of refraction for the fibers of 1.6, the nominal speed of light is calculated to be $18.73$ cm/ns or $0.625c$. The experimentally determined speeds of light may not match the nominal case due to the fact that photons do not travel down the fiber in a straight line, but rather travel helically down the fiber, increasing their effective path length and thus lowering the measured speed of light found using the straight-line path assumption.

\subsubsection{Position measurement from timing}
\label{sec:position_from_timing}

\begin{figure}[h!]
	\centering
	\begin{minipage}[t]{.52\textwidth}
		\centering
		\includegraphics[width=\textwidth]{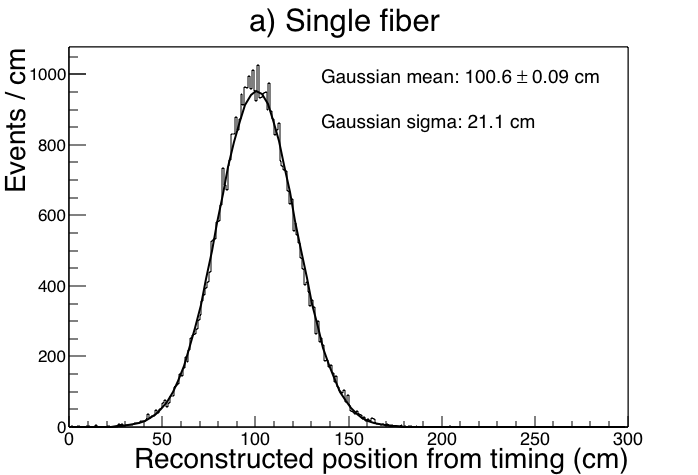}
			
	\end{minipage}%
	\hspace*{0.01cm}
	\begin{minipage}[t]{.52\textwidth}
		\centering
		\includegraphics[width=\textwidth]{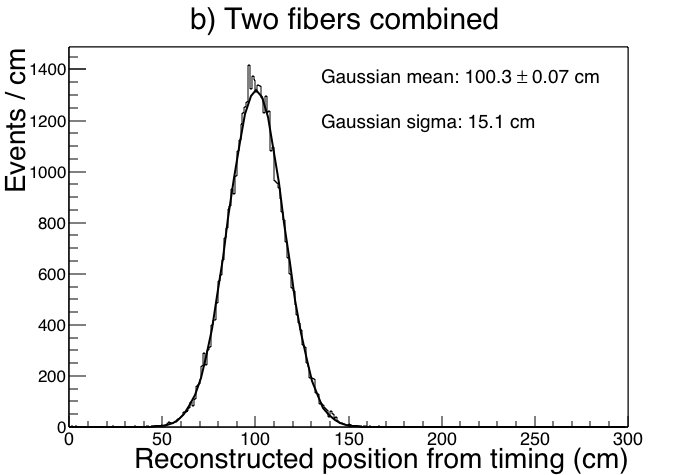}
		
	\end{minipage}
	\caption{The average reconstructed position of proton traversals using light arrival time differences from the two ends of a fiber when the beam was positioned at a nominal 1\,m from both ends of the dicounter and centered in the top counter.  Calculations were done with a single fiber (a) and averaging both fibers in a counter (b).  The beam position is well reconstructed with a resolution of about 0.15\,m when two fibers are combined.}
	\label{fig:position_from_timing}
\end{figure}

Using the speed of light in the fibers and the light arrival time difference, the incident position of the proton along the counter length can be computed using Eq.~\ref{eqn:x_position}. Figure~\ref{fig:position_from_timing}(a) shows the distribution of reconstructed positions using timing information from both ends of a single fiber for a run where the beam was incident $1$\,m from the ends of a dicounter of type C.  A fit was applied using a Gaussian function giving an average constructed position of $100.6 \pm 0.1$\,cm and a standard deviation of $21.1$\,cm using a single fiber.  The longitudinal positions of the counters were found to be misaligned by up to $8$\,mm, so the reconstruction position is in good agreement with the true position.  By averaging the time difference in two fibers, the position resolution is improved by a factor of $1/\sqrt{2}$ to about $15$\,cm, see Fig.~\ref{fig:position_from_timing}(b)\footnote{The position may also be estimated by comparing the light yield response at each end of a counter, but the resolution from that method is found to be approximately a factor of two worse than the resolution using timing.}.

\begin{figure}[h!]
	\centering
        \includegraphics[width=0.6\textwidth]{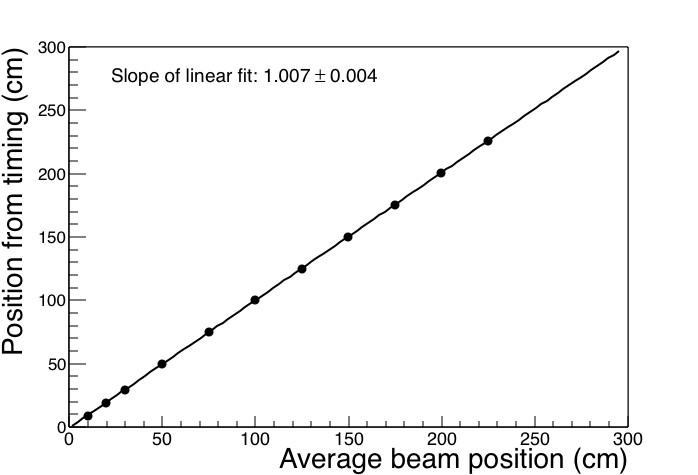}
	\caption{The average reconstructed position obtained from combining the timing information in two fibers is shown as a function of the average incident proton position as calculated from the MWPC data. Note that the errors on the positions are taken as the error on the mean from the Gaussian fits (see Fig.~\ref{fig:position_from_timing}) and are too small to be visible on the figure.}
	\label{fig:positions}
\end{figure}

Proton positions were reconstructed based on timing and compared to the average proton traversal position from the MWPC data when the beam was centered transversely in a counter and incident at various positions along the longitudinal length of the dicounter.   The average reconstructed event position as a function of average incident proton position is shown in Fig.~\ref{fig:positions}.  Applying a linear fit to the data in Fig.~\ref{fig:positions} gave a slope consistent with 1.0 indicating that the incident proton position could be accurately reconstructed.  

\section{Summary and Conclusions}

A pulsed beam of $120~\mathrm{GeV}$ protons provided at the Fermilab Test Beam Facility was used to measure properties of the prototype counters for the Cosmic Ray Veto system of the Mu2e experiment.  The counters were constructed using extruded scintillation strips with titanium dioxide coating and embedded wavelength-shifting fibers read out by silicon photomultipliers.

The average PE yield was measured to be $50~\mathrm{PE}$ per SiPM channel for normal incident protons at a position $1~\mathrm{m}$ from the end of a $3~\mathrm{m}$-long counter read out with 1.4\,mm fibers and $2.0{\times}2.0$\,mm$^2$ SiPMs.  With increased concentration of TiO$_2$ in the co-extruded counter coating, the average PE yield increased by more than 30\% to over $65~\mathrm{PE}$ per SiPM channel.  Studies of the impact of changing the fiber size, and the amount of dopant were also performed.

Longitudinal and transverse beam scans were used to study properties of the prototype counters and the light yield from angular scans was found to be consistent with our expectations according to the length of the  path taken through the counter.  Single-channel timing resolution based on a 79.5~MHz sampling rate was demonstrated to be better than 2~ns. With a single-channel timing resolution of less than $2~\mathrm{ns}$, the particle position can be determined using the time difference from readouts at each end of the fiber to 15\,cm by combining timing information from both fibers in a counter.

The Mu2e counters will be outfitted with 1.4\,mm diameter WLS fibers, the most recent vintage $2.0{\times}2.0$\,mm$^2$ Hamamatsu SiPMs, and will be extruded with the improved TiO$_2$  concentration in the co-extruded coating.  With this configuration, the Cosmic Ray Veto system will meet the requirements of the Mu2e experiment.

\section{Acknowledgements}

We are grateful for the vital contributions of the Fermilab staff and the technical staff of the participating institutions.
This work was supported by the US Department of Energy; 
the Italian Istituto Nazionale di Fisica Nucleare;
the Science and Technology Facilities Council, UK;
the Ministry of Education and Science of the Russian Federation;
the US National Science Foundation; 
the Thousand Talents Plan of China;
the Helmholtz Association of Germany;
and the EU Horizon 2020 Research and Innovation Program under the Marie Sklodowska-Curie Grant Agreement No.690385. 
Fermilab is operated by the Fermi Research Alliance, LLC under Contract No. DE-AC02-07CH11359 with the U.S. Department of Energy, Office of Science, Office of High Energy Physics. The United States Government retains and the publisher, by accepting the article for publication, acknowledges that the United States Government retains a non-exclusive, paid-up, irrevocable, world-wide license to publish or reproduce the published form of this manuscript, or allow others to do so, for United States Government purposes.


\end{document}